\documentclass{acta}
\usepackage{supertabular,lscape,epsfig}
\usepackage{amssymb}
\usepackage{amsmath}
\usepackage{multirow}
\usepackage{float}
\usepackage{journals}
\usepackage{textcomp}
\usepackage{tabularx}
\mathchardef\mhyphen="2D
\usepackage{placeins}
\DeclareSymbolFont{ppa}{OT1}{ppl}{m}{it}
\DeclareMathSymbol{\vv}{\mathalpha}{ppa}{'166}

\SetPages{0}{999}

\SetVol{999}{2025}

\usepackage{csquotes}
\usepackage{xpatch}
\usepackage[maxalphanames=1,labelalpha,style=authoryear,language=english,citestyle=authoryear-comp,maxcitenames=2,mincitenames=1,natbib=true,backend=biber,sorting=anyt,bibencoding=utf8,maxbibnames=2,minbibnames=2,labeldateparts=true,uniquelist=false,uniquename=false,sortcites=ya,giveninits=true]{biblatex}
\DeclareNameAlias{sortname}{family-given}

\newcolumntype{C}[1]{>{\centering\arraybackslash}p{#1}}
\DefineBibliographyStrings{english}{andothers = {\em et\addabbrvspace al\adddot}}
\xpatchbibmacro{name:andothers}{%
  \bibstring[\emph]{andothers}%
}{%
  \bibstring[\emph]{andothers}%
}{}{}

\DeclareBibliographyDriver{article}{
\usebibmacro{begentry}
    \refitem
        {\printnames{author}}
        {\printfield{year}}
        {\thefield{journaltitle}}
        {\printfield{volume}}
        {\printfield{pages}}
    \usebibmacro{finentry}
}

\newcommand{\TabCapp}[2]{\begin{center}\parbox[t]{#1}{\centerline{
  \small {\spaceskip 2pt plus 1pt minus 1pt T a b l e}
  \refstepcounter{table}\thetable}
  \vskip2mm
  \centerline{\footnotesize #2}}
  \vskip3mm
\end{center}}

\newcommand{\MakeTableee}[4]{\begin{table}[htb]\TabCapp{#2}{#3}
  \begin{center} \TableFont \begin{tabular}{#1} #4
  \end{tabular}\end{center}\end{table}}

\newfont{\bb}{ptmbi8t at 12pt}
\newfont{\bbb}{cmbxti10}
\newfont{\bbbb}{cmbxti10 at 9pt}

\usepackage{geometry}
\begin{document}
\begin{Titlepage}
\Title{The OGLE Collection of Variable Stars.\\
Over 75\,000 Eclipsing and Ellipsoidal Binary Systems in the Magellanic Clouds\footnote{Based on observations obtained with the 1.3-m Warsaw telescope at the Las Campanas Observatory of the Carnegie Institution for Science.}}
\Author{
M.~~G~ł~o~w~a~c~k~i$^1$,~~
I.~~S~o~s~z~y~ń~s~k~i$^1$,~~
A.~~U~d~a~l~s~k~i$^1$,~~
M.\,K.~~S~z~y~m~a~ń~s~k~i$^1$,~~
J.~~S~k~o~w~r~o~n$^1$,\\
D.\,M.~~S~k~o~w~r~o~n$^1$,~~
P.~~M~r~ó~z$^1$,~~
P.~~P~i~e~t~r~u~k~o~w~i~c~z$^1$,~~
R.~~P~o~l~e~s~k~i$^1$,~~
S.~~K~o~z~ł~o~w~s~k~i$^1$,\\
P.~~I~w~a~n~e~k$^1$,~~
M.~~W~r~o~n~a$^{2,1}$,~~
K.~~U~l~a~c~z~y~k$^{3,1}$,~~
K.~~R~y~b~i~c~k~i$^{4,1}$,\\
M.~~G~r~o~m~a~d~z~k~i$^1$~~
M.~~M~r~ó~z$^1$,~~
and~~~M.~~U~r~b~a~n~o~w~i~c~z$^1$
}
{$^1$Astronomical Observatory, University of Warsaw, Al.~Ujazdowskie~4,\\ 00-478~Warszawa, Poland\\
$^2$Department of Astrophysics and Planetary Science, Villanova University, 800 East Lancaster Avenue, Villanova, PA 19085, USA\\
$^3$Department of Physics, University of Warwick, Gibbet Hill Road,
Coventry, CV4~7AL,~UK\\
$^4$Department of Particle Physics and Astrophysics, Weizmann Institute of Science, Rehovot 76100, Israel}
\Received{March 999, 2025}
\end{Titlepage}

\Abstract{We present an updated collection of eclipsing and ellipsoidal binary systems in the Large and Small Magellanic Clouds (LMC and SMC), as observed by the Optical Gravitational Lensing Experiment (OGLE) survey. The catalog comprises a total of $75\,400$ binary systems, including $63\,252$ in the LMC and $12\,148$ in the SMC. The sample is categorized into $67\,971$ eclipsing and $7429$ ellipsoidal variables. For all stars, we provide {\it I}-band and {\it V}-band photometric time series collected between 2010 and 2024 during the fourth phase of the OGLE project (OGLE-IV). We discuss methods used to identify binary systems in the OGLE data and present objects of particular interest, including double periodic variables, transient eclipsing binaries, double eclipsing binaries, and binary systems with pulsating stars. We present a comparative analysis based on the most comprehensive catalogs of variable stars in the Magellanic System, compiled from surveys like Gaia, ASAS-SN, and EROS-2, and included in the International Variable Star Index.}{binaries: eclipsing, ellipsoidal - Catalogs}
\setcounter{page}{1}
\Section{Introduction}
More than 50$\%$ of all stars are part of binary or multiple star systems \parencite{sana, duchene}. Binaries are a crucial source of data for modern astrophysics, as they enable direct measurement of fundamental stellar parameters, including the masses, radii, surface temperatures, and absolute magnitudes of their components \parencite{andersen1991}. Additionally, detached eclipsing binaries offer the ability to determine intergalactic distances with better than 1$\%$ accuracy \parencite{pietrzynski2019} and provide a unique opportunity to carry out stringent tests of our understanding of stellar structure and evolution. Photometric variations in binary systems can arise from several mechanisms, including eclipses, tidal ellipsoidal distortions, reflection effects, beaming effects, and a variety of phenomena occurring in accretion disks.

The bulge and disk of the Milky Way, along with the Magellanic Clouds, provide natural environments for extensive studies of binary systems. The Optical Gravitational Lensing Experiment (OGLE) project published catalogs containing over half a million eclipsing and ellipsoidal binary systems, including more than $450\;000$ systems in the Galactic bulge \parencite{soszynski2016blgecl}, over $11\;000$ in the Galactic disk \parencite{pietrukowicz2013}, approximately $40\;000$ in the Large Magellanic Cloud (LMC), and more than $8000$ in the Small Magellanic Cloud (SMC) \parencite{pawlak2016ecl}. These numbers can be compared to the catalogs of binary systems released by other large-scale sky surveys: $154\;600$ objects in the ASAS-SN catalog of variable stars \parencite{asas}, $34\;562$ objects in the EROS-2 database \parencite{epoch}, and over 2~million eclipsing binary candidates published as a part of the {\it Gaia} Data Release 3 \parencite{gaiaECL}. Currently, The International Variable Star Index (VSX; \cite{vsx}), which compiles variable objects from literature, includes more than one million eclipsing and ellipsoidal variables\footnote{\texttt{https://www.aavso.org/vsx/} accessed 11.12.2024}.

Catalogs of variable stars are invaluable in many branches of astronomy. Previously released OGLE catalogs of binary systems enabled search for exoplanets as presented by \citet{mrozPlanety} or search for binary system candidates hosting a black hole or neutron star \parencite{kapusta}. The collection was used by various authors to, among others, study fundamental stellar parameters and tidal interactions \parencite{tides}, to train neural networks for automatic classification of variable stars \parencite{nn}, to study the apsidal motion of eccentric binaries \parencite{apsidal}, to analyze double eclipsing systems \parencite{doublyecl} and to study light travel time effect \parencite{lighttravel}.

In this paper, we present an updated collection of eclipsing and ellipsoidal binaries in the Magellanic System. The primary motivation for updating the previous edition of the OGLE catalog of binary systems in the Magellanic Clouds \parencite{pawlak2016ecl} is to include newly identified objects and to refine existing classifications. Several additional years of OGLE observations collected after the publication of the \citet{pawlak2016ecl} catalog allowed us to identify $27\;578$ new eclipsing and ellipsoidal variables, to reclassify and remove 783 objects from the previous version of the collection, to refine the orbital periods, magnitudes, amplitudes, and other parameters of the systems, and finally to detect long-term changes and secondary periodicities in numerous variables.

\Section{Observations and Data Reductions}
The photometric data used in this study were collected during the OGLE-IV project, between March 2010 and May 2024, with the 1.3-meter Warsaw telescope at Las Campanas Observatory in Chile (the observatory is operated by the Carnegie Institution for Science). This telescope is equipped with a mosaic camera composed of 32 CCDs, each with 2048 by 4102 pixels, providing a total field of view of 1.4 square degrees. The surveyed area of the Magellanic Clouds region includes 546 OGLE-IV fields, covering roughly 765 square degrees.

Observations were conducted using the Cousins {\it I}-band and  the Johnson {\it V}-band filters. The vast majority of the data ($\sim$90\%) were collected using the {\it I}-band filter with an exposure time of 150~s. The remaining {\it V}-band observations were secured with the same integration time. The {\it I}-band magnitude range captured by OGLE-IV extends from 13~mag to 21.5~mag. The number of epochs collected reaches 5808 in the LMC and 2489 in the SMC, with a median number of epochs equal to 879 and 2274 respectively. The minimal number of epochs collected is 14 in both LMC and SMC.

Photometry was carried out using the Difference Image Analysis method \parencite{DIA1, DIA2}. Procedure for the calibration to the standard photometric system together with the detailed information on the OGLE-IV instrumental setup and data reduction system have been described by \citet{ogle4}.

\Section{Selection and Classification of Binary Systems}
Initially, the {\it I}-band time-series photometry of 70 million point sources observed by OGLE in the LMC and 14 million in the SMC was processed using the {\sc Fnpeaks}\footnote{http://helas.astro.uni.wroc.pl/deliverables.php?active=fnpeaks\&lang=en} program to search for periodic variability. The probed frequency space ranged from 0 to 24~cycles per day with a resolution of $10^{-5}$~cycles per day. Subsequently, two additional period search algorithms were applied to the light curves with the highest signal-to-noise ratios for the initially derived periods: the string-length method \parencite{dworetsky} and the traditional least-squares fitting of a fourth-order Fourier series. The string-length method is particularly well-suited for highly non-sinusoidal light curves, such as those of eclipsing stars with narrow eclipses. In contrast, the Fourier fitting method is more effective at determining periods for light curves that exhibit continuous variations, as observed in contact and semi-detached binaries. Detected frequencies were refined using the {\sc Tatry} code based on multi-harmonic periodogram \parencite{czerny} within $\pm10\%$ range.

The next step of our procedure involved the visual inspection of the light curves phase-folded with the identified periods. After excluding previously known variable stars, we focused on light curves with a signal-to-noise ratio of the detected periodicity greater than 4. To maintain the accuracy of the catalog, candidates for new eclipsing and ellipsoidal variables were independently reviewed by two researchers. In this way, more than $20\;000$ eclipsing and ellipsoidal variables, not included in the previous version of the OGLE collection \parencite{pawlak2016ecl}, were identified. This sample was further expanded with several thousand binaries found in the OGLE \parencite{poleski2010dpv, wrona2022} and external catalogs of variable stars, including VSX \parencite{vsx}, EROS-2 \parencite{epoch}, ASAS-SN \parencite{asas}, and Gaia DR3 \parencite{gaiaECL}. More details on comparison between catalogs are provided in section \ref{crossmatch} Ultimately, the OGLE collection of variable stars was extended by $27\,578$ previously overlooked binary systems in the Magellanic Clouds.

The photometry of binary systems included in the \citet{pawlak2016ecl} catalog was updated with new OGLE observations, and the orbital periods, magnitudes, amplitudes, and other observational parameters of these variables were rederived. After visually inspecting these light curves, we decided to remove 783 objects (1.6\% of the total sample) from the catalog, as we reclassified them as pulsating stars, spotted variables, or other types of variable stars. Additionally, for approximately 3,700 binaries in the \citet{pawlak2016ecl} catalog, we significantly (by more than 1\%) modified their orbital periods.

In the final step of the procedure for selecting and classifying binary systems, we divided our sample into candidates for contact and non-contact eclipsing variables, as well as ellipsoidal variables. The latter group includes eccentric ellipsoidal variables (aka heartbeat stars; \cite{wrona2022}). Since our classification is based solely on the features of light curves, it should be treated with caution.

\Section{Binary Systems in the Magellanic Clouds}
The OGLE collection of binary systems in the Magellanic Clouds consists of $75\;400$ objects ($63\;252$ in the LMC and $12\;148$ in the SMC), of which $67\;971$ are eclipsing stars and 7421 are ellipsoidal variables. The data for all objects can be accessed through the OGLE web interface or FTP sites:
\begin{center}
{\it https://ogle.astrouw.edu.pl $\rightarrow$ OGLE Collection of Variable Stars}\\
{\it https://www.astrouw.edu.pl/ogle/ogle4/OCVS/lmc/ecl/}\\
{\it https://www.astrouw.edu.pl/ogle/ogle4/OCVS/smc/ecl/}\\
\end{center}

The identifiers of eclipsing binaries follow the convention established in the catalogs of \citet{graczyk2011ecl} and \citet{pawlak2016ecl}. Each variable is labeled using the format OGLE-LMC-ECL-NNNNNN or OGLE-SMC-ECL-NNNNNN (where NNNNNN is a six-digit number), based on whether it was located in the sky region covering the LMC or SMC. The boundary between these regions is defined at a celestial meridian of $2.8^{\text{h}}$. Ellipsoidal variables in our catalog are designated according to the format OGLE-LMC-ELL-NNNNNN or OGLE-SMC-ELL-NNNNNN.

\begin{figure}
    \centering
    \includegraphics[width=\linewidth]{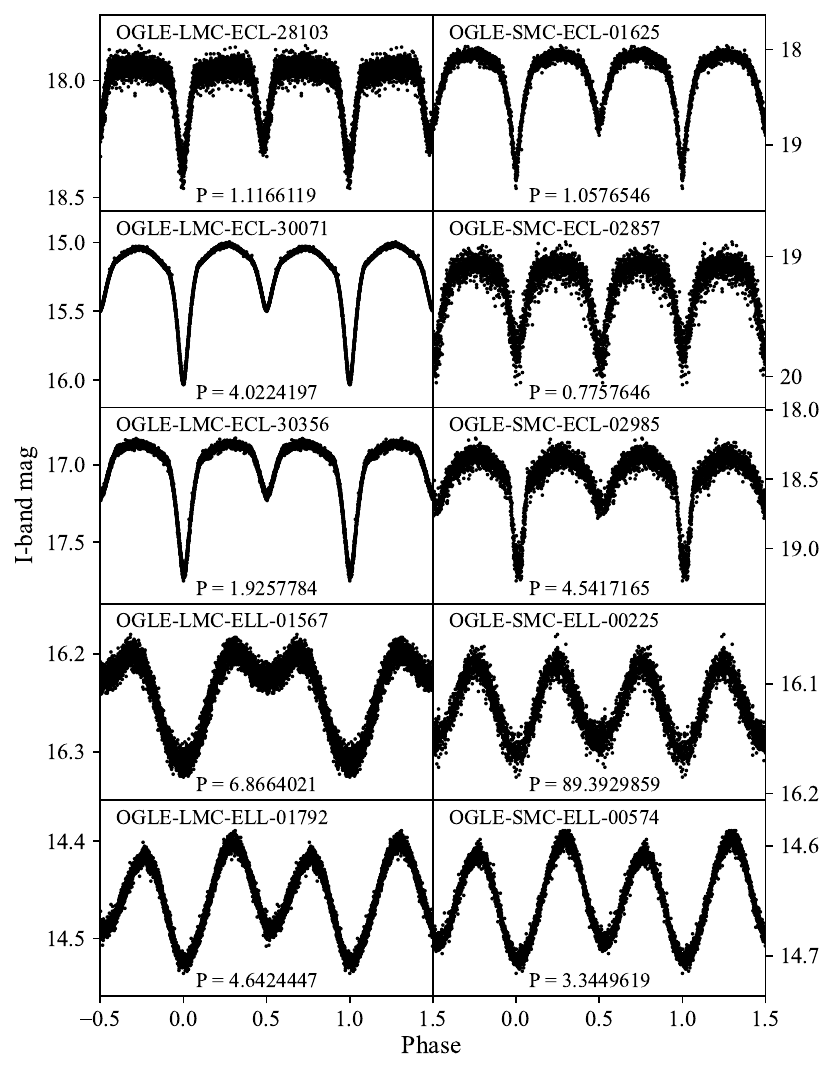}
    \caption{Examples of light curves of eclipsing and ellipsoidal variables included in this collection. Left column contains objects from the LMC, right one from the SMC. Upper three rows show eclipsing binaries, while bottom two ellipsoidal variables.}
    \label{fig:eg-lc}
\end{figure}

Observational parameters of all variables, such as equatorial coordinates, orbital periods, out-of-eclipses {\it I}-band and {\it V}-band magnitudes, depths of the primary and secondary eclipses, and epochs of the primary eclipse are provided. Depths of the eclipses were measured by finding the baseline and local minima of the light curves smoothed with median filter and Savitzky-Golay filter \parencite{savgol}. Time series {\it I}-band and {\it V}-band photometry obtained during the OGLE-IV survey (2010-2024) is also available. This photometry may be combined with the light curves from previous phases of the OGLE survey \parencite{udalski1998ecl, wyrzyk2003ecl, wyrzyk2004ecl, graczyk2011ecl, pawlak2013ecl, pawlak2016ecl}, providing three decades long photometric time series, perfect for studying variability in long time scales. However, offsets may occur between the photometric zero points, due to different filters and CCD detectors used during particular stages of the project, as well as due to blending and crowding. Only OGLE-IV light curves were used to determine orbital period and other parameters included in the catalog, unless a given object was observed only during previous surveys. Examples of light curves included in the collection are presented in Fig.~\ref{fig:eg-lc}.

The file remarks.txt contains a list of stars with uncertain classification, along with variables exhibiting atypical properties, \eg binary systems with a pulsating component, double periodic variables, RS CVn stars, transient eclipsing binaries, etc.

The completeness of our collection of eclipsing and ellipsoidal
variables depends on many factors, such as their position in the sky,
brightness, amplitude of light variations, period, shapes of their light
curves, etc. The highest completeness is expected in the central regions
of the LMC and SMC, which have been monitored since 1997 by the OGLE-II
and subsequently by the OGLE-III and OGLE-IV projects, and searched for
eclipsing variables by \citet{udalski1998ecl}, \citet{wyrzyk2003ecl, wyrzyk2004ecl}, \citet{graczyk2011ecl} and \citet{pawlak2013ecl, pawlak2016ecl}. Naturally, our
sample does not include objects brighter than the saturation level of OGLE
photometry ($I\approx13$~mag) and stars too faint for their variability
to be detected and classified ($I\gtrsim20$~mag).

The formal completeness of the catalog was assessed using eclipsing and
ellipsoidal variables found in overlapping regions of adjacent OGLE-IV
fields. These stars could be independently detected twice, although, in
the final version of our collection each object is represented by only a
one entry from the OGLE database, typically the one with a light curve
containing more data points. Considering only light curves with 50 or
more data points, we {\it a posteriori} found 6018 stars with double
identifications in the OGLE-IV database, so we had a chance to find
$12\;036$ counterparts. We independently identified 7886 of them,
whereas in 2102 cases, only one component of the pair was identified.
Consequently, this leads to the catalog completeness level of
approximately 79\%. As expected, the faintest stars were the most
difficult to identify, along with those exhibiting the smallest
amplitudes, detached systems with very long periods, and eclipsing
variables displaying additional variability.

\begin{figure}[h]
    \centering
    \includegraphics[width=\linewidth]{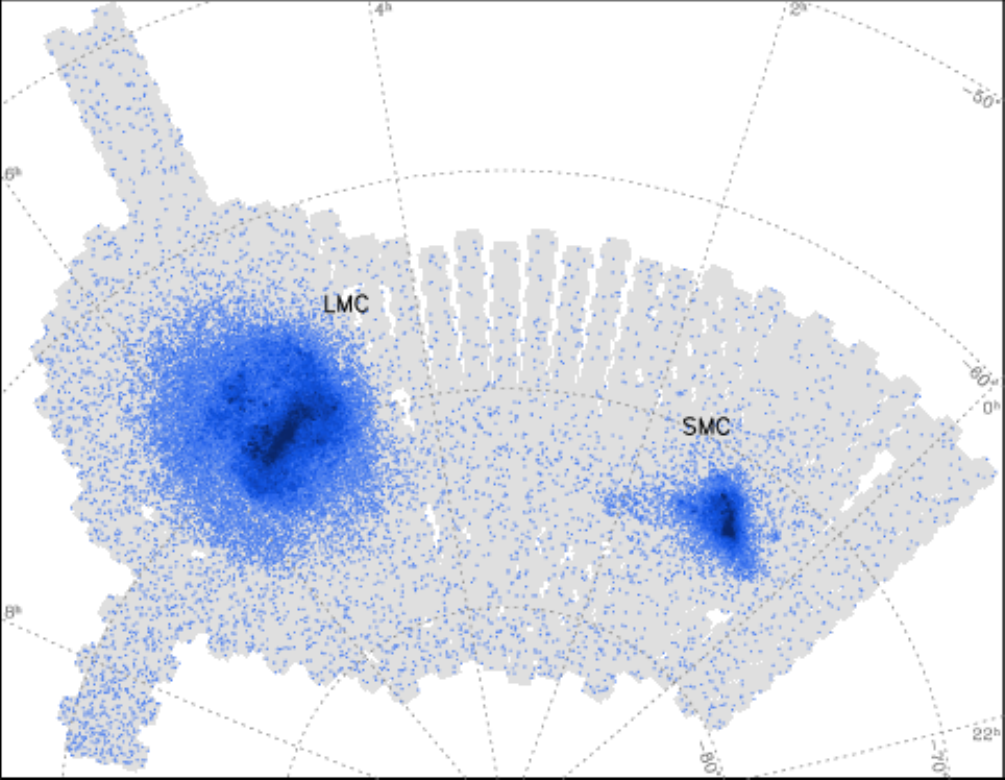}
    \caption{Spatial distribution of eclipsing and ellipsoidal variables from the OGLE catalog located in the Magellanic Clouds. Darker color of the points indicates higher density of stars in given region. Light gray color in the background represents OGLE-IV fields in the Magellanic System.}
    \label{fig:positions}
\end{figure}

\Section{Discussion}
\Subsection{Catalog statistics}
A simple statistical analysis was performed on the basic parameters of the cataloged objects. Their positions are shown in Fig.~\ref{fig:positions}, which demonstrates that the vast majority of eclipsing and ellipsoidal variables are members of the LMC and SMC. However, some stars lie in the foreground of both galaxies and belong to the Milky Way. Based on the spatial distribution of the objects in the outer fields of the OGLE-IV footprint, we estimate that our collection includes approximately $5000$ binary systems from our Galaxy.

\begin{figure}
    \centering
    \includegraphics[width=\linewidth]{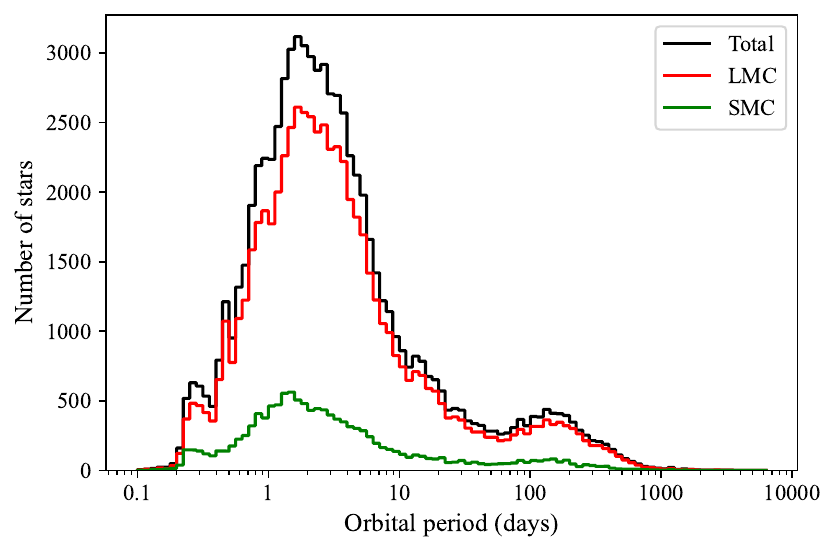}
    \includegraphics[width=\linewidth]{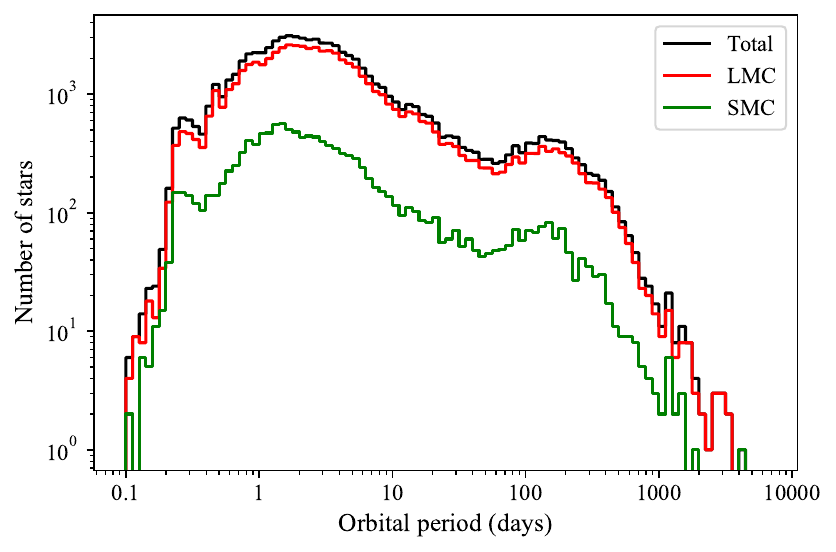}
    \caption{Distribution of orbital periods of eclipsing and ellipsoidal variables in the Magellanic Clouds. The number of stars is presented in the linear (upper panel) and logarithmic scale (lower panel). The red, green, and black lines represent histograms for binaries in the LMC, SMC, and both Clouds combined, respectively.}
    \label{fig:periods}
\end{figure}

Fig.~\ref{fig:periods} shows the histograms of orbital periods of binary systems in the Magellanic Clouds. Both LMC and SMC binaries follow similar period distributions, which are also consistent with the distribution of periods obtained by \citet{soszynski2016blgecl} for eclipsing and ellipsoidal binaries in the Galactic bulge. In particular, the period distributions of binary systems in both the Magellanic Clouds and the Milky Way exhibit a local maximum for periods between 100 and 200 days, which is mainly caused by close systems (usually ellipsoidal variables) consisting of red giant stars. In contrast, the main peak of the period distribution for binaries in the Magellanic System occurs at around 2 days, which does not align with the maximum at 0.4 days for stars observed toward the Galactic bulge \parencite{soszynski2016blgecl}. This discrepancy arises from a selection effect: most short-period contact systems (W UMa stars) in the Magellanic Clouds are too faint to be detected with the 1.3-meter OGLE telescope. Indeed, objects classified as contact binary systems constitute only 4\% of our sample of eclipsing variables in the Magellanic Clouds (a significant portion of them lies in front of the Magellanic Clouds and belongs to the Milky Way halo), whereas in the Galactic bulge, more than 20\% of all eclipsing stars are W~UMa-type systems.

\begin{figure}
    \centering
    \includegraphics[width=0.93\linewidth]{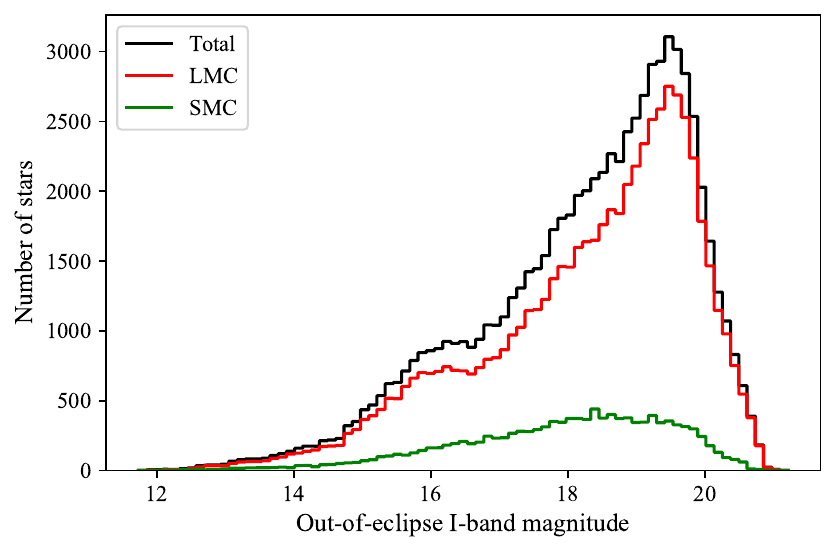}
    \caption{Distribution of out-of-eclipse {\it I}-band magnitudes of eclipsing and ellipsoidal variables from the OGLE catalog located in the Magellanic Clouds.}
    \label{fig:mags}
\end{figure}

The distribution of out-of-eclipse {\it I}-band magnitudes of our variables is shown in Fig.~\ref{fig:mags}. The shape of this distribution reflects both the luminosity function of binary stars in the Magellanic Clouds and the detection limits for variable stars using OGLE photometry. The histogram has a maximum around $I=19.5$~mag. For fainter stars, the completeness of our catalog declines sharply. Additionally, the LMC distribution has another maximum at around 16~mag, most probably due to a high population of luminous stars from the tip of the red-giant branch.

\begin{figure}
    \centering
    \includegraphics[width=0.93\linewidth]{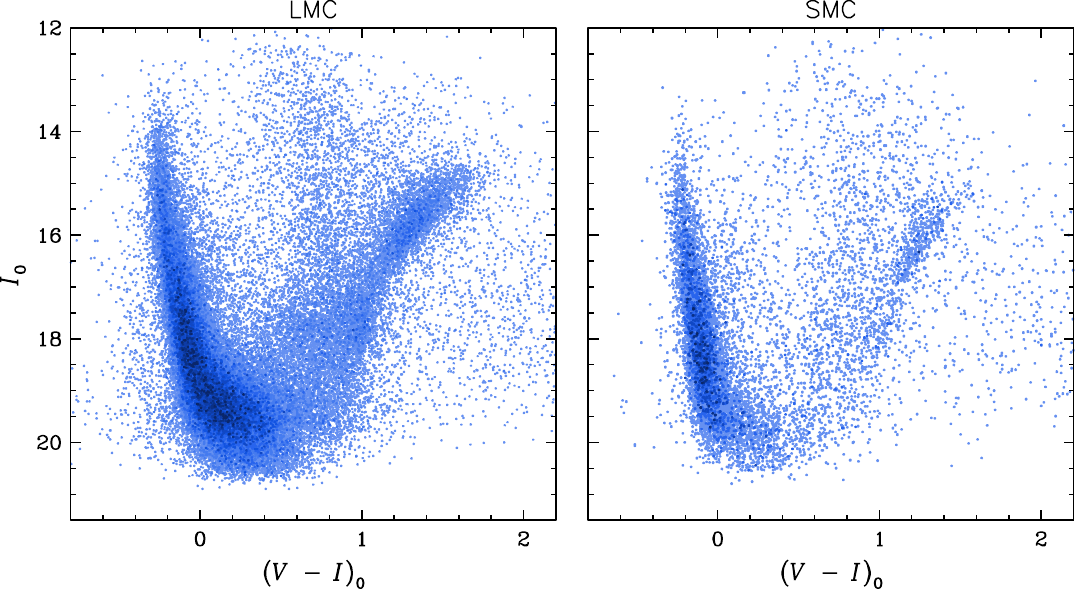}
    \caption{Color-magnitude diagram for eclipsing and ellipsoidal variables from the OGLE catalog located in the Magellanic Clouds. The color of points indicates the density of stars.}
    \label{fig:cmd}
\end{figure}

All LMC and SMC objects were presented in a color-magnitude diagram (Fig.~\ref{fig:cmd}), after correcting their $V-I$ color index and {\it I}-band magnitude for extinction using reddening maps created by \citet{skowron2021}. The diagram shows that roughly 75\% of the stars in our catalog lie on the main sequence, while the majority of the remaining systems contain red giant~stars.


\Subsection{Cross-match with other catalogs}
\label{crossmatch}
As mentioned in Section 3, to enhance the completeness of our collection, we cross-matched point sources observed by OGLE with candidate eclipsing binaries published in the ASAS-SN \parencite{asas}, VSX \parencite{vsx}, EROS-2 \parencite{epoch}, and Gaia DR3\parencite{gaiaECL} catalogs. A careful visual inspection of the OGLE light curves of the stars selected in this way allowed us to increase the size of our collection by approximately 5000 objects.
Some candidates from other catalogs were rejected because of lack of periodic variability in the OGLE data or because the variability did not match with binary systems. The remaining objects were also excluded from the final version of this collection because of too small amplitude, too high brightness, inaccurate coordinates in different catalogs, localization in the area corresponding to the technical gaps between CCDs, etc.

\begin{figure}[H]
    \centering
    \includegraphics{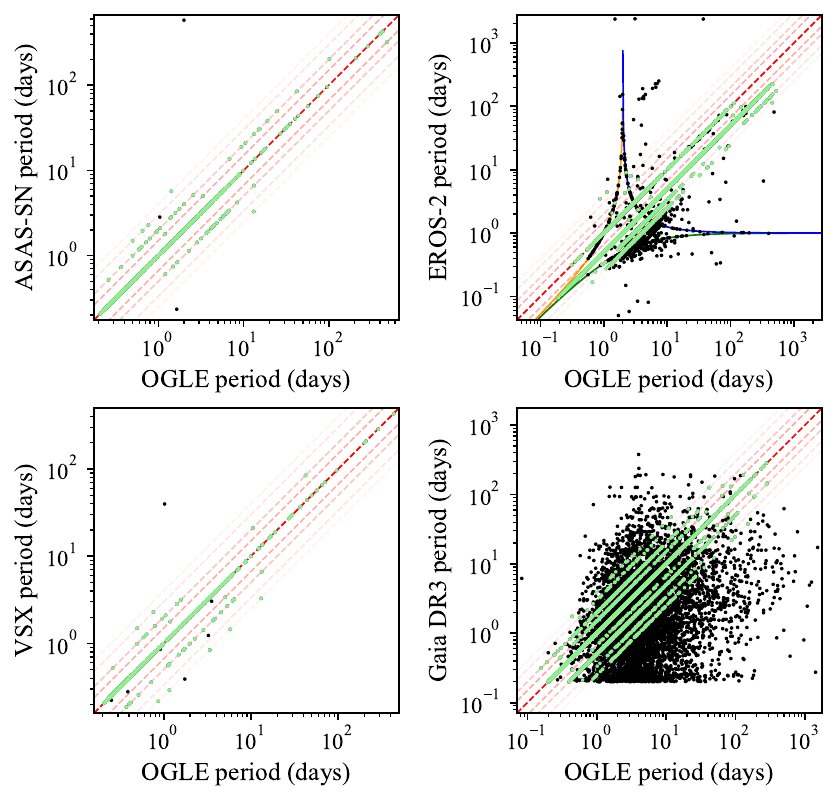}
    \caption{Comparison of the orbital periods measured with OGLE data with the orbital periods from external catalogs. The red dashed lines represent orbital period harmonics defined in the equation \ref{eq:multiples}, with the intensity of the color decreasing for higher values of $k$. The green points represent objects with matching periods.}
    \label{fig:period_com}
\end{figure}

For each external catalog, orbital periods of all matched objects were compared with the main orbital periods from the presented catalog and with their harmonics as follows:
\begin{equation}
    P_{k} = k^{\pm 1} P_{\rm OGLE} \quad \text{for} \quad k \in \left\{1, \frac{3}{2}, 2, 3, 4 \right\}
\label{eq:multiples}
\end{equation}

where $P_{\rm OGLE}$ is the orbital period measured from the OGLE light curves and provided in our collection. Figure.~\ref{fig:period_com} compares the orbital periods for ASAS-SN, VSX, EROS-2 and \textit{Gaia} DR3 respectively. The summary results of the cross-matching with other catalogs are presented in Table~1.

\MakeTableee{|c|c|c|c|c|}{1pt}{Comparison of binary stars catalogs}{
        \hline
        Catalog & ASAS-SN & VSX & EROS-2 & \textit{Gaia} DR3 \\ \hline
        Objects within OGLE fields in the MC & 1707 & 1675 & 53 942 & 38 827 \\ \hline
        Matched with OGLE catalog & 65\% & 62\% & 43\% & 61\% \\ \hline 
        Objects misclassified as binaries & 7.5\% & 0.66\% & 5.4\% & 0.82\% \\ \hline 
        Objects rejected from the OGLE catalog & 5\% & 10\% & 50\% & 35\% \\ \hline 
        Matching period & 95\% & 96\% & 4.1\% & 65\% \\ \hline 
        Matching period (with multiples) & 99.7\% & 99.3\% & 97.3\% & 75.3\% \\ \hline
}

For ASAS-SN and VSX catalogs, almost all objects have orbital periods directly matching with our data. However, the orbital periods provided in the EROS-2 catalog differ significantly from the values measured from the OGLE light curves. Most of them are equal to half the expected values. The discrepancies between the EROS-2 data and the OGLE data were also described by \citet{poleski2014rr}. Additionally, the following period aliases are also included in EROS-2 data (see upper right panel in Fig.~\ref{fig:period_com}):
\begin{flalign}
f_{1} &= 2f_{\rm OGLE} + 1 &\text{(dark green line)} && \label{eq:first}\\
f_{2} &= 2f_{\rm OGLE} - 1 &\text{(orange line)} && \label{eq:second}\\
f_{3} &= -2f_{\rm OGLE} + 1 &\text{(blue line)} && \label{eq:third} 
\end{flalign} where $f_{i}=P_{i}^{-1}$. There is a very significant discrepancy between \textit{Gaia} DR3 and OGLE orbital periods, which does not follow any regular pattern. This suggests high degree of randomness in orbital periods in \textit{Gaia} DR3 catalog, which also means that those stars were correctly classified as eclipsing or ellipsoidal variables by coincidence.

\Subsection{Binaries with other types of variability}
Among binary stars included in this work, there are interesting groups of objects showing additional variability. The selected groups are described below.

\begin{figure}
    \centering
    \includegraphics{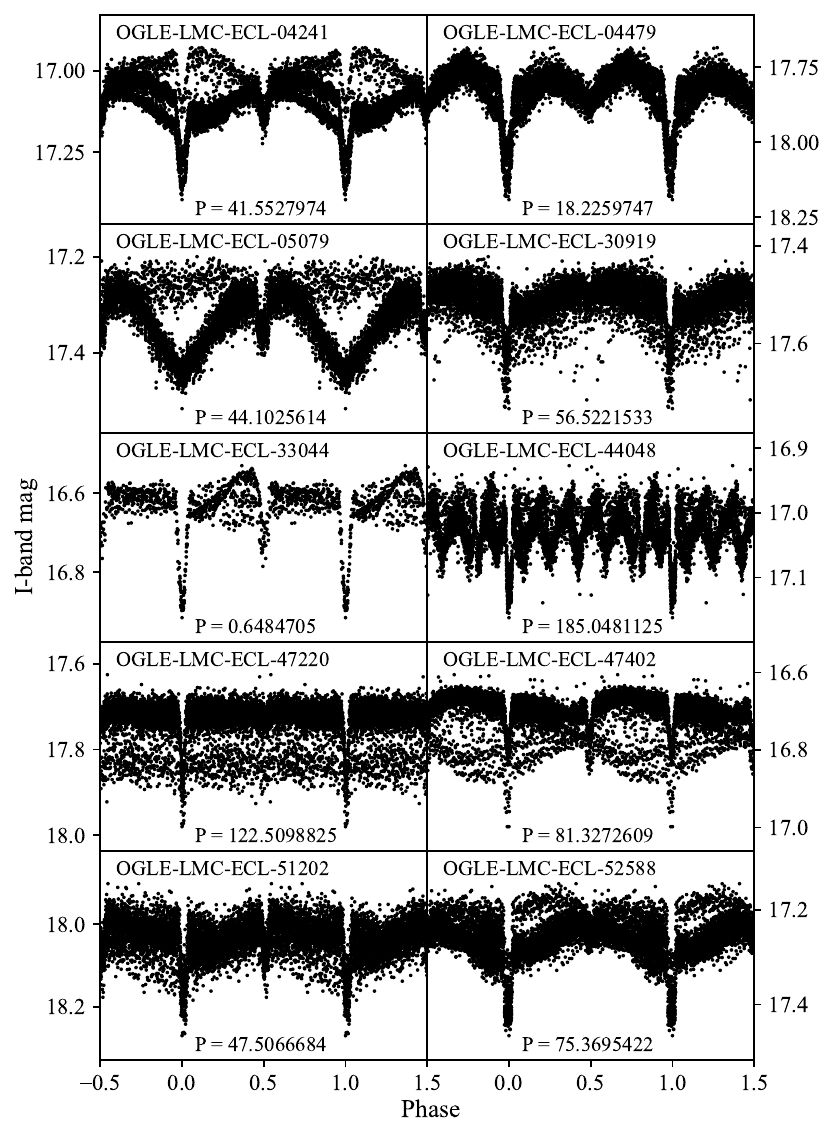}
    \caption{Light curves of selected RS CVn stars folded with the orbital periods.}
    \label{fig:rscvn}
\end{figure}

\hfill \\
{\bf RS Canum Venaticorum} (RS CVn) stars are binary objects with darker spots on the surface, introducing additional semi-regular variability between eclipses. The light curves of selected objects are presented in Fig.~\ref{fig:rscvn}. Our collection contains at least 294 variables of this type.

\begin{figure}
    \centering
    \includegraphics{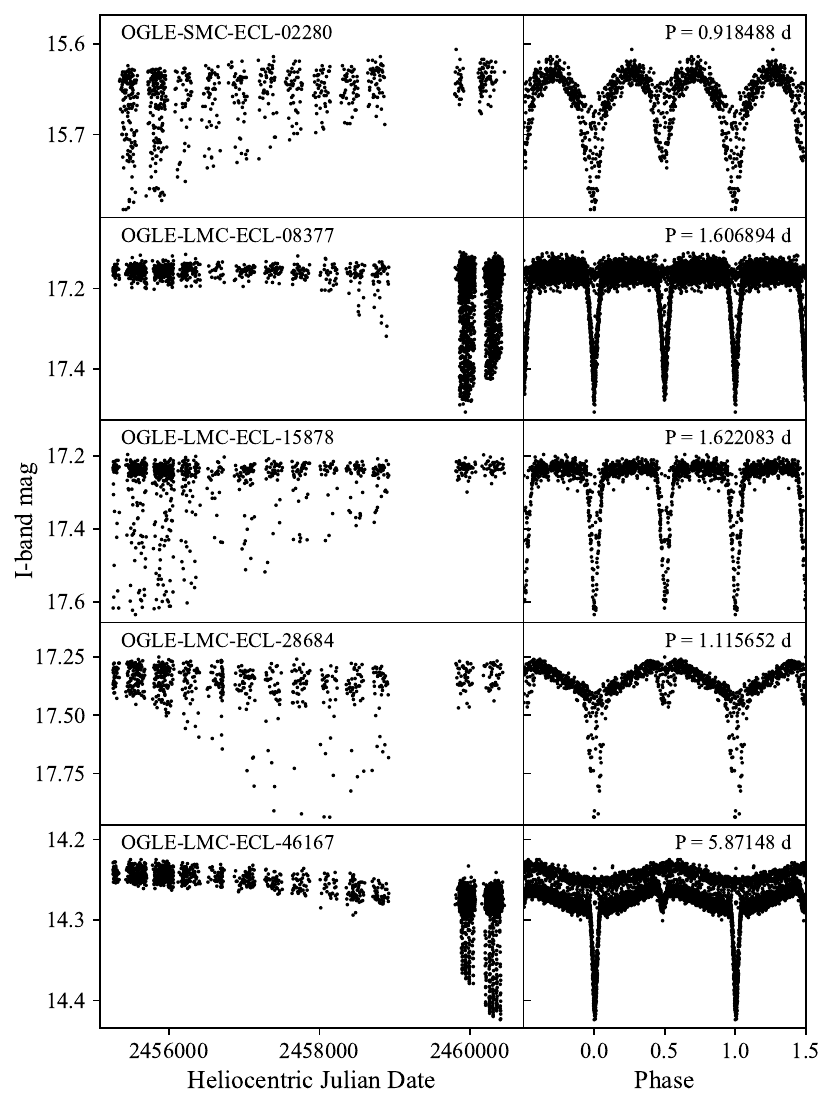}
    \caption{Phase-folded light curves of selected transient eclipsing binaries from this work. Each row represents one object, showing both time-domain and phase-folded light curve.}
    \label{fig:transient}
\end{figure}

\begin{figure}
    \centering
    \includegraphics[width=0.8\linewidth]{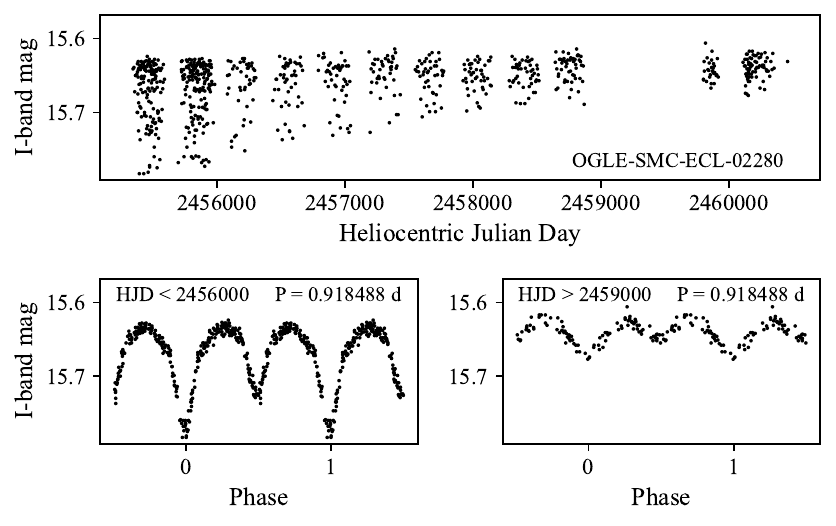}
    \caption{Transient eclipsing object retaining ellipsoidal variability when eclipsing behavior ceases. The upper panel shows unfolded light curve obtained from OGLE-IV data. The left lower panel shows phase-folded light curve from first three observational seasons (with visible eclipses), whereas the right panel shows the phase-folded data from the last two seasons (with ellipsoidal variability).}
    \label{fig:ellips}
\end{figure}

\begin{figure}
    \centering
    \includegraphics[width=0.8\linewidth]{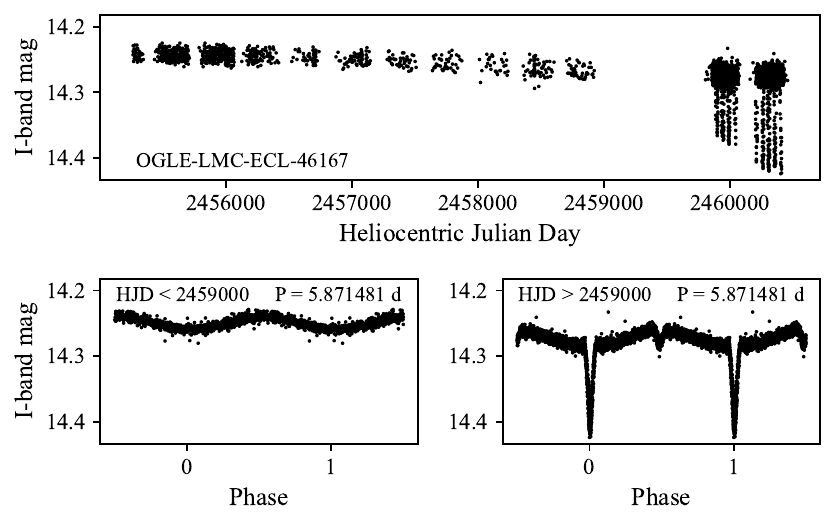}
    \caption{Transient eclipsing object showing reflection effect with later appearing eclipses. The upper panel shows unfolded light curve obtained from OGLE-IV data. The left lower panel was detrended for clarity due to change of average magnitude over time, and shows phase-folded light curve from first 11 observational seasons with visible reflection effect. The right panel shows eclipsing behavior of the object in the last two seasons.}
    \label{fig:refl}
\end{figure}

\hfill \\
{\bf Transient eclipsing binaries} - these objects do not show eclipsing variations at all times. For example, eclipses may gradually appear or disappear, or may occur in cycles. They were first described by \citet{graczyk2011ecl}. Analysis of such a long term variability is possible thanks to combined OGLE photometric data from multiple years of observations. Some of the transient eclipsing binaries from this catalog are presented in Fig.~\ref{fig:transient}. When these stars do not show eclipsing variability, they may retain ellipsoidal behavior (Fig.~\ref{fig:ellips}), or reflection effect (Fig.~\ref{fig:refl}).


\begin{figure}
    \centering
    \includegraphics{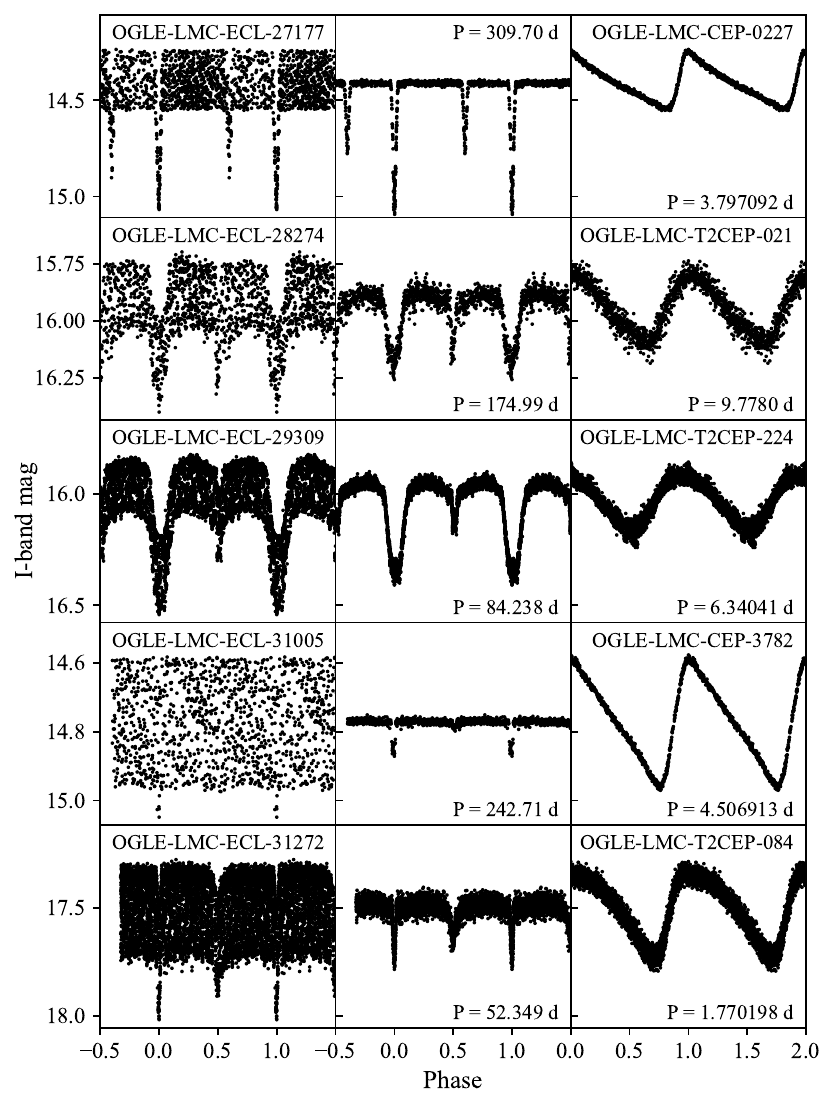}
    \caption{Light curves of selected systems with pulsating stars from this work. Left panels show the original light curves folded with the orbital periods, while middle and right panels present the disentangled eclipsing and pulsating light curves, respectively.}
    \label{fig:cap}
\end{figure}

\hfill \\
{\bf Binary systems with pulsating stars} $-$ binary systems where at least one of the components is a pulsating star (e.g. a Cepheid or RR Lyrae star). Fig.~\ref{fig:cap} contains some of such objects from this collection.

\begin{figure}
    \centering
    \includegraphics{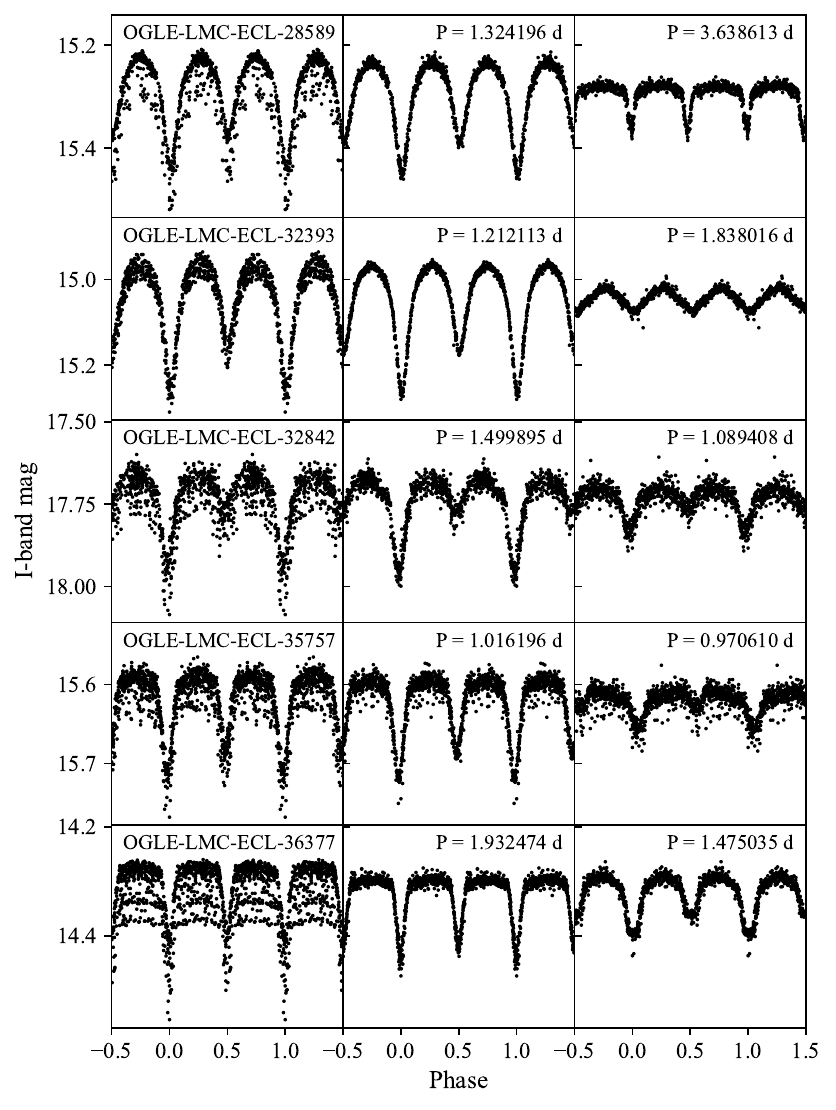}
    \caption{Light curves of selected double eclipsing systems from this work. Left panels display the original light curves folded with one of the orbital periods, while middle and right panels present the disentangled light curves folded with both orbital periods.}
    \label{fig:var}
\end{figure}

\hfill\\
{\bf Double eclipsing binaries} are objects in which two superimposed eclipsing or ellipsoidal modulations are simultaneously visible. Our collection contains more than 150 such systems. The light curves for selected stars are presented in Fig.~\ref{fig:var}. Most probably they are quadruple systems consisting of two pairs of eclipsing/ellipsoidal variables orbiting common center of mass.

\begin{figure}
    \centering
    \includegraphics{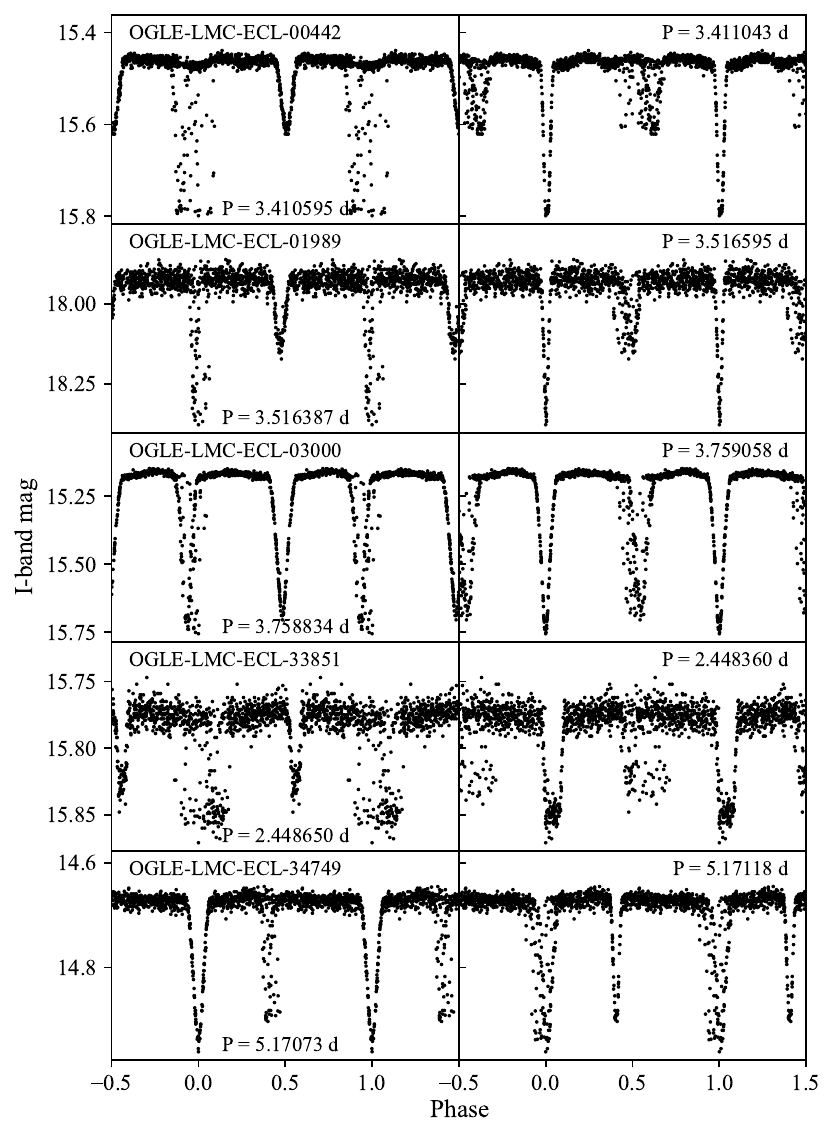}
    \caption{Phase-folded light curves of selected systems with strong apsidal motion from this work. Each row represents the same object, with light curve phase-folded with slightly different periods.}
    \label{fig:apsidal}
\end{figure}

\hfill \\
{\bf Systems with apsidal motion} $-$ due to a strong apsidal motion, each eclipse for these objects phases with slightly different period. Examples of such objects from this collection are presented in Fig.~\ref{fig:apsidal}.

\begin{figure}
    \centering
    \includegraphics{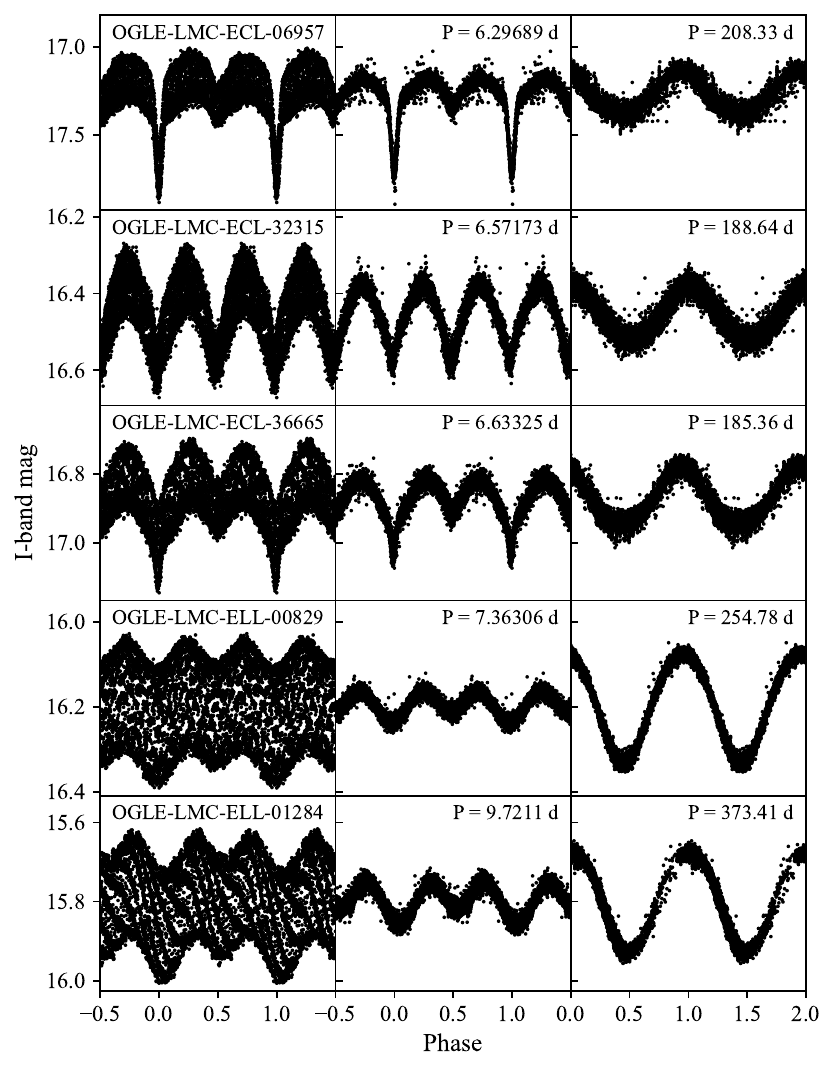}
    \caption{Light curves of selected DPV stars from this work. Left panels show the original light curves folded with the orbital periods, while middle and right panels present the disentangled eclipsing and long periodic variability light curves, respectively.}
    \label{fig:dpvs}
\end{figure}

\begin{figure}
    \centering
    \includegraphics{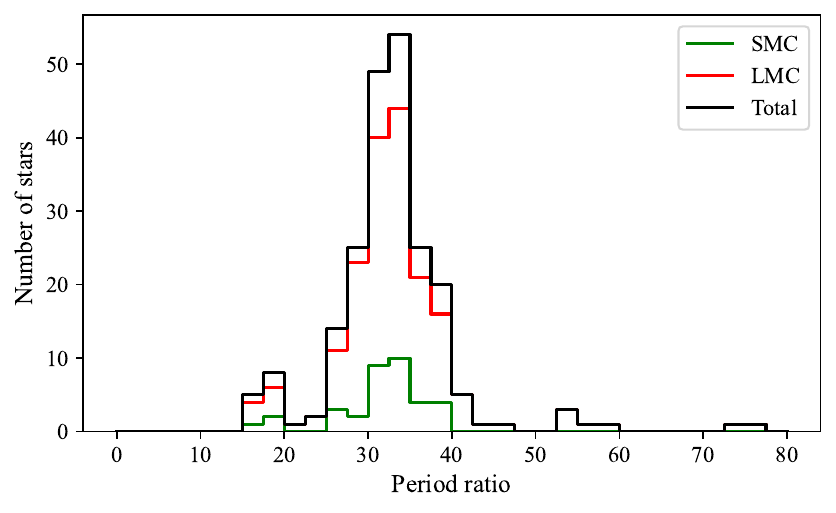}
    \caption{Distribution of ratios of longer periods to orbital periods for DPV stars included in this catalog.}
    \label{fig:dpv_hist}
\end{figure}

\hfill \\
{\bf Double periodic variables} (DPVs) are binary stars with a long secondary period, first recognized by \citet{mennickent2003dpv}. Typically, it is around 33 times greater than the orbital period. This catalog contains 220 objects of this type, extending the previous OGLE catalog of DPVs published by \citet{poleski2010dpv} by 95 new objects. The selected objects are presented in Fig.~\ref{fig:dpvs}. The distribution of the ratios of their longer periods to shorter periods is shown in Fig.~\ref{fig:dpv_hist}.


\begin{figure}[h]
    \centering
    \includegraphics{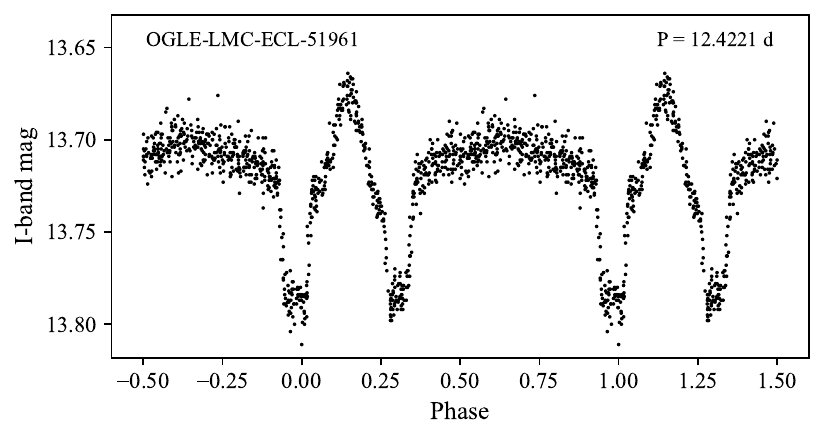}
    \caption{Phase-folded light curve of a peculiar binary object. The object becomes significantly brighter between eclipses, which separation is also quite low. Most probably it is eccentric ellipsoidal variable with visible eclipses.}
    \label{fig:pec1}
\end{figure}

\begin{figure}[h]
    \centering
    \includegraphics{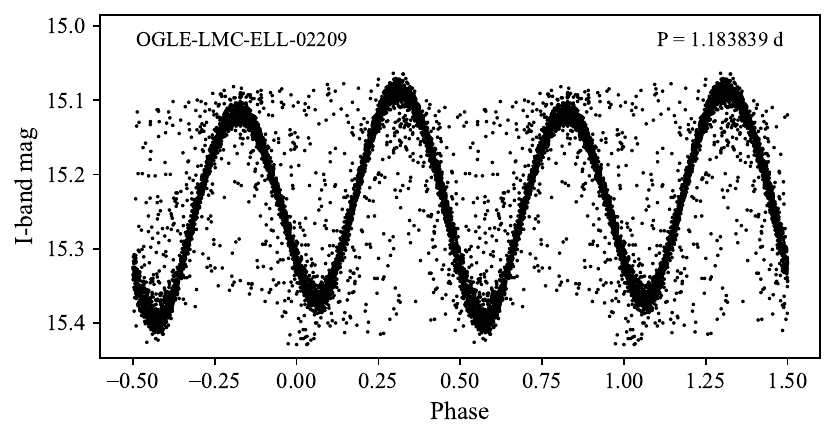}
    \caption{Phase-folded light curve of ellipsoidal variable with significant changes of orbital period. Its light curve was divided into year-long segments and phase-folded with different orbital periods in Fig. \ref{fig:pec4}.}
    \label{fig:pec2}
\end{figure}

\begin{figure}[h]
    \centering
    \includegraphics{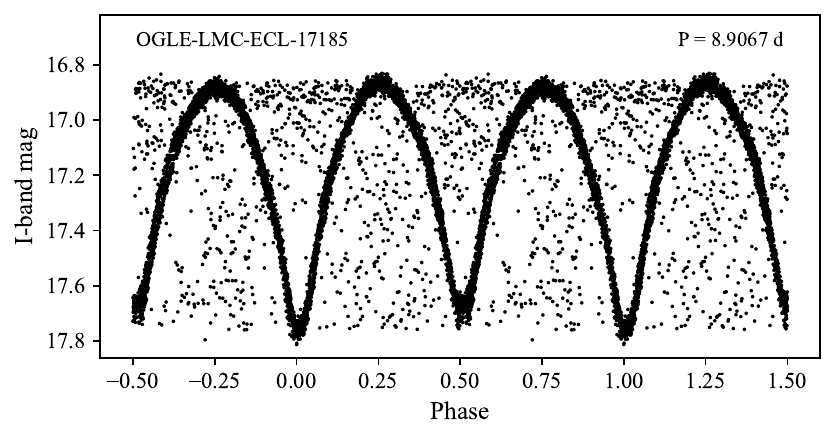}
    \caption{Phase-folded light curve of eclipsing variable with significant changes of orbital period.Its light curve was divided into year-long segments and phase-folded with different orbital periods in Fig. \ref{fig:pec5}.}
    \label{fig:pec3}
\end{figure}

\begin{figure}[H]
    \centering
    \includegraphics{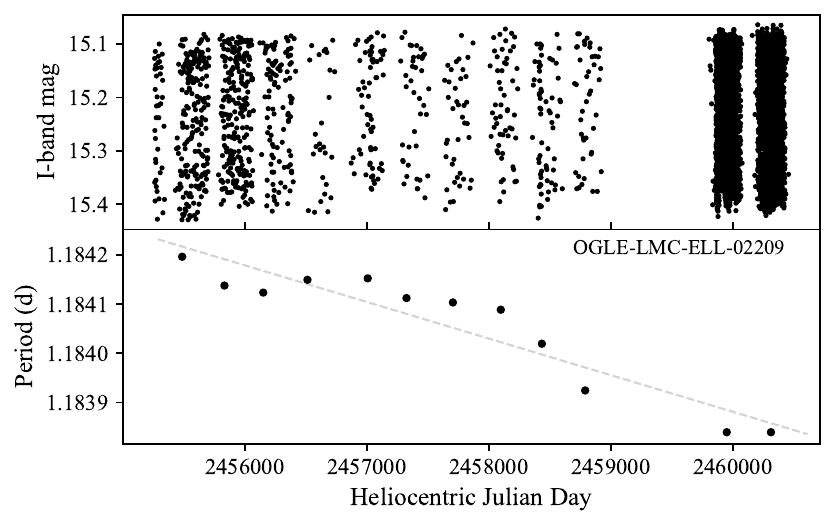}
    \includegraphics{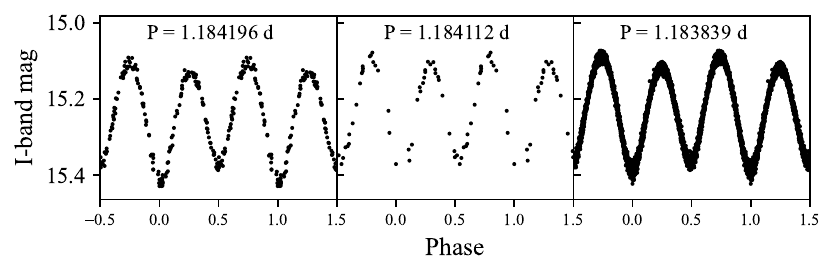}
    \caption{Upper panel shows the light curve and manually adjusted orbital period for year-long fragments. The gray dashed line represents the fitted linear function, used to determine the rate of period change: $2.72\times10^{-5}$~d~yr$^{-1}$. The lower panel shows pieces of this light curve phase-folded from the first season, season in the middle, and the last season with manually adjusted periods. Despite changes of the orbital period, the shape of the light curve remains unchanged.}
    \label{fig:pec4}
\end{figure}

\begin{figure}[H]
    \centering
    \includegraphics{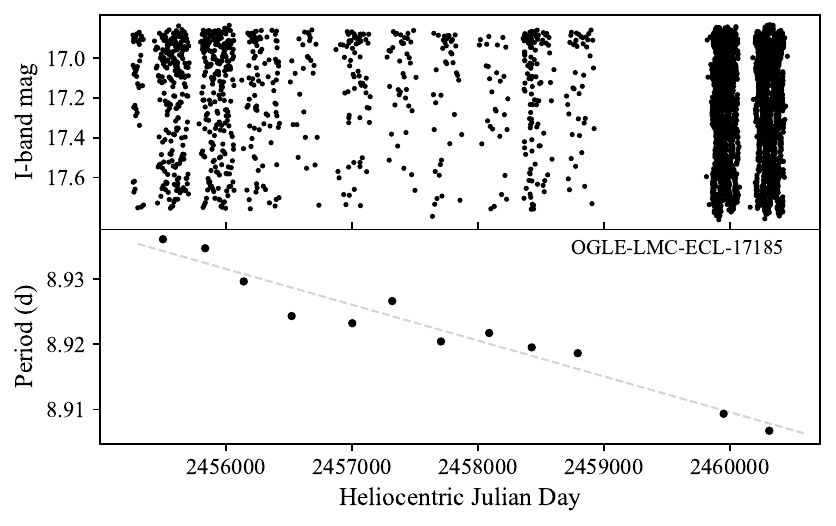}
    \includegraphics{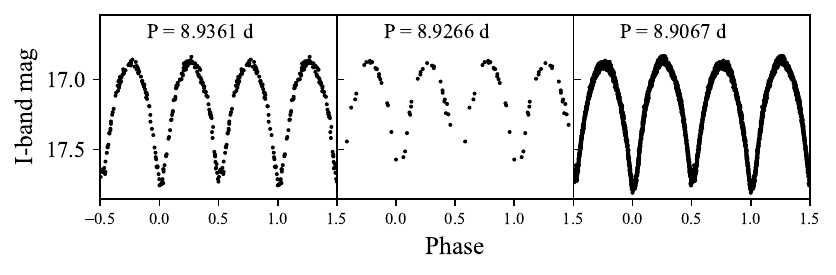}
    \caption{Upper panel shows the light curve and manually adjusted orbital period for year-long fragments. The gray dashed line represents the fitted linear function, used to determine the rate of period change: $2.01\times10^{-3}$~d~yr$^{-1}$. The lower panel shows pieces of this light curve phase-folded from the first season, season in the middle, and the last season with manually adjusted periods. In the last observational period the object started exhibiting the O'Connell effect.}
    \label{fig:pec5}
\end{figure}

\hfill\\
{\bf Peculiar objects} $-$ some objects exhibit additional variability that is difficult to classify and explain, for example, the OGLE-LMC-ECL-51961 object (presented in Fig. \ref{fig:pec1}). Most probably it is an eccentric ellipsoidal variable with $\omega\approx0^\circ$ or $\omega\approx180^\circ$ as described in \citet{wrona2022}, with visible eclipses. Other examples of interesting objects are OGLE-LMC-ELL-02209 and OGLE-LMC-ECL-17185 (presented in Figs. \ref{fig:pec2} and \ref{fig:pec3} respectively) showing significant decrease in orbital period over time. Their light curves were divided into year-long segments, and each segment was phase-folded with manually adjusted orbital period (Figs. \ref{fig:pec4} \& \ref{fig:pec5}). The linear function was fitted to determine the average rate of period changes: $2.72\times10^{-5}$~d~yr$^{-1}$ and $2.01\times10^{-3}$~d~yr$^{-1}$ respectively. The latter object also started exhibiting the O'Connell effect in the last observational season. Analogous behavior was observed in the red nova V1309 Scorpii progenitor before its outburst \parencite{rednova}, with an average period decrease of $3.46\times10^{-3}$~d~yr$^{-1}$ and similar changes in the shape of the phase-folded light curve.

\clearpage

\clearpage
\Section{Conclusions}
We presented the OGLE collection of eclipsing and ellipsoidal variables in the Magellanic Clouds. The new catalog contains over 50\% more objects than the one previously published by \citet{pawlak2016ecl}, and it is the largest catalog of such binary systems in the Magellanic Clouds up to date. For each object we provide orbital period, depth of eclipses, out-of-eclipse magnitude, color, and long time-series {\it I}-band and {\it V}-band photometry.

Our collection contains unusual objects of particular interest: double periodic variables, transient eclipsing binaries, binaries with pulsating stars, chromospherically active stars, systems with strong apsidal motion, double eclipsing binaries, and many other bizarre variables. 

\Acknow{This work has been funded by the National Science Centre, Poland, grant no.~2022/45/B/ST9/00243. For the purpose of Open Access, the author has applied a CC-BY public copyright license to any Author Accepted Manuscript (AAM) version arising from this submission.}

\end{document}